\documentclass{article}

\usepackage{arxiv}

\usepackage[utf8]{inputenc} 
\usepackage[T1]{fontenc}    
\usepackage{hyperref}       
\usepackage{url}            
\usepackage{booktabs}       
\usepackage{amsfonts}       
\usepackage{nicefrac}       
\usepackage{microtype}      
\usepackage{lipsum}
\usepackage{graphicx}
\usepackage{caption}
\usepackage{subcaption}
\graphicspath{ {./images/} }

\title{SysMap: A Lightweight Software Visualization Tool to Analyze the Software Evolution of a System}

\author{
 Fazle Rabbi \\
  Institute of Information Technology\\
  University of Dhaka\\
  Dhaka, Bangladesh \\
  \texttt{bsse0725@iit.du.ac.bd} \\
   \And
Nishat Tasnim Niloy \\
  Institute of Information Technology\\
  University of Dhaka\\
  Dhaka, Bangladesh \\
  \texttt{bsse0725@iit.du.ac.bd} \\
  \And
Nadia Nahar \\
  Institute of Information Technology\\
  University of Dhaka\\
  Dhaka, Bangladesh \\
  \texttt{nadia@iit.du.ac.bd} \\
  \And
Md. Nurul Ahad Tawhid	 \\
  Institute of Information Technology\\
  University of Dhaka\\
  Dhaka, Bangladesh \\
  \texttt{tawhid@iit.du.ac.bd} \\
}

\begin{document}
\maketitle
\begin{abstract}
Software visualization helps to comprehend the system by providing a vivid illustration. The developers, as well as the analysts, can have a glance over the total system to understand the basic changes over time from a high-level point of view through this technique. In recent years, many tools are proposed to visualize software based on different architectural metaphors, such as as- solar system, city or park. Some of the solutions have just worked on system visualization where a few tried to explain the changes in software throughout different versions that still need heavy manual work. Keeping such limitations in mind, this paper proposes a lightweight tool named SysMap that takes the source codes of different versions of software systems, provides 3D illustrations of those systems and a graphical statistic of its evolution. To build the graphical element to represent the system, the source code has been studied to find out different software metrics. For experimentation, several open-source java projects were chosen to find out the necessary information. Henceforth, this tool will surely increase the work efficiency of both the developer and analyst by reducing the manual effort and by providing the graphical view to comprehend the software evolution over time. An video demonstration of that tool is attached in a web url\footnote{https://drive.google.com/file/d/17WeEuPoVgedmrCmm4uns-ImRyDv0sDJV/view?usp=sharing}. 
\end{abstract}

\keywords{Software Visualization \and Software Evolution \and Architectural Metaphors \and Software Analysis \and Software Metrics.}

\section{Introduction}
\label{sec:introduction}
Maintaining software is a must to meet up the ever-changing demands of the users. This requires the analyst to understand the ins and outs of the system thoroughly. Evolution is the way of maintaining, changing or updating the software for the inevitable changing requirements \cite{bennett2000software}. Software visualization is one of the most convenient way to get a quick estimation of a system no matter how complex it is \cite{ball1996software}, \cite{andrews1997information}, \cite{charters2002visualisation} and \cite{charters2002visualisation}. For more than two decades 2D and 3D visualization approaches have been used extensively. However, 3D visualization in Software Engineering is still a new area to be studied.

In this paper, A 3D visualization tool (SysMap) is proposed where an object-oriented software system can easily be visualized, explored and analyzed. SysMap can show packages and classes in a software project as plots and buildings respectively. The metaphors are based on the calculated metrics from the source codes of the systems. Previously, \cite{alam2007evospaces}, \cite{balogh2013codemetropolis}, \cite{balogh2015codemetropolis} have proposed some tools which shows illustrations of very high resolutions. Even game engines are used to make the interactions interesting. However, SysMap intends to provide very lightweight approach to ensure a smooth or easy interaction from user end . Hence,  a minimal graphical representation and basic metrics calculations are used.

The proposed system is implemented as a tool named SysMap. This tool can handle the large open-source Java projects like - Proguard, Jaimlib as well as the smaller student projects. The user can explore the map, interact with the elements of the map and analyze the buildings to get an overview of the software. The elements can provide the information of eight metrics - line of code (LOC), comment percentage, the coupling between objects, lack of cohesion, weighted method per class (WMC), number of children and inheritance level.

Some solutions have already worked with the visualization for system comprehension. However, to the best of the authors' knowledge, no tool has provided such a version to version system illustration features that help to visualize all the versions at once and provides a graph of the software evolution. In Section \ref{sec:res}, the evolution of five versions of Proguard is shown graphically. 

The contributions of this paper are:
\begin{itemize}
    \item A 3D visualization approach that provides the illustration of the elements of a software system in a minimal fashion.
    \item A description and demonstration of a tool named SysMap which helps to visualize and navigate both the commercial large systems and student projects. A video demonstration of the tool is also provided in a video in this Google drive link \cite{drive}. 
    \item Simultaneous visual representation of different versions of a software system.
    \item A graphical analysis of the evolution of different versions of a system on the basis of their changing metrics values.
\end{itemize}

In Section \ref{sec:meth}, the methodology of the proposed tool is provided. In Section \ref{sec:tool} the description of the tool-SysMap is elaborated. The output of the tool is discussed in Section \ref{sec:res} and in Section \ref{sec:threat}, the threats to validity are mentioned. Finally, the related works are presented in Section \ref{sec:rel} and conclusion in Section \ref{sec:con}.    


\section{Methodology}
\label{sec:meth}
\begin{figure}[ht]
  \centering
  \includegraphics[width=.5\textwidth]{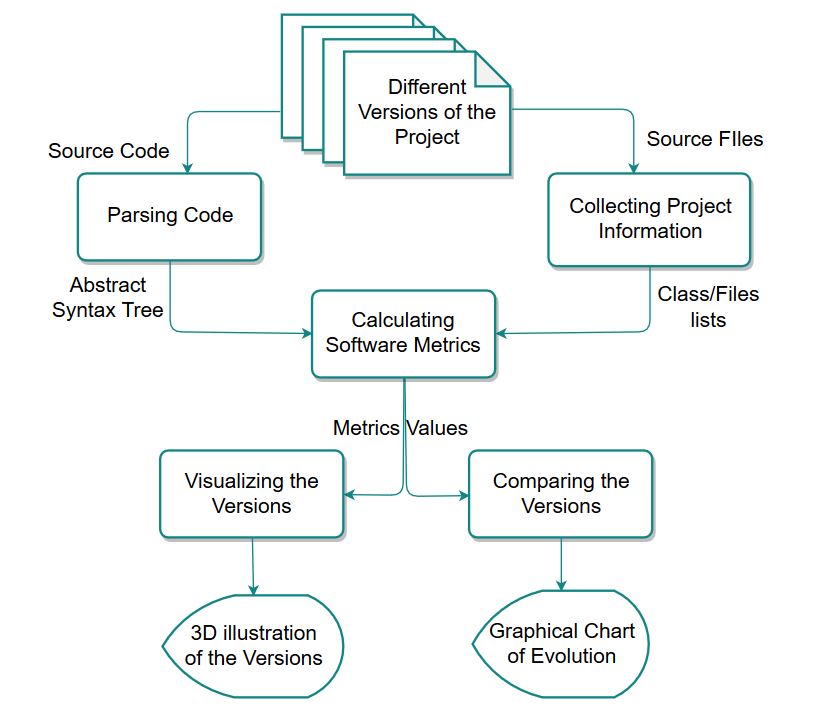}  
  \caption{Flowchart of the SysMap System}
  \label{fig:flowchart}
\end{figure}

The proposed approach has been divided into four subsystems - data collection, metrics analysis, software visualization and evolution analysis. Figure \ref{fig:flowchart} depicts the full work flow of the tool. The subsystems are discussed in the following subsections:

\begin{table}[ht]
\centering
\caption{Information about the Collected Java Projects}
\begin{tabular}{|c|c|c|c|c|}
\hline
\multicolumn{1}{c|}{\textbf{Project Type}} & \textbf{Software} & \textbf{Package} & \textbf{Class} & \textbf{LOC} \\ \hline
Open Source    & Proguard          & 37              & 782             & 40750   \\ \hline
 Open Source    & JDeodorant-master & 22              & 376             & 59463   \\ \hline
Open Source    & Jaimlib           & 04              & 046             & 01054  \\ \hline
Student Project   & Account Management          & 01        & 10              & 633   \\ \hline
Student Project   & AES Encryption          & 01        & 09              & 301   \\ \hline  
Student  Project  & SysMap                      & 10        & 21              & 1795   \\ \hline
\end{tabular}
\label{tab:input}
\end{table}

\subsection{Data Collection}
\label{sec:datacoll}
For the dataset to visualize the system, three open-source java projects have been collected named- Proguard, JDeodorant-master, and Jaimlib. Two student lab projects of the undergraduate level were collected named Account Management and AES Implementation. and the source code of SysMap itself. Table \ref{tab:input} shows information regarding the java projects as input. Both the commercial and the student projects have been taken into consideration to visually understand the differences between the code architecture types. To analyze the evolution of a software, different versions of same systems have been collected. Some versions of the previously mentioned open source projects were available. However, for the student projects, none of those had any second releases. In Table \ref{tab:proguard}, the collected information about different releases of Proguard is shown.

\begin{table}[ht]
\centering
\caption{Information about Different Versions of Proguard}
\begin{tabular}{|c|c|c|c|c|}
\hline
\textbf{No} & \textbf{Version} & \textbf{Package} & \textbf{Class} & \textbf{LOC} \\ \hline
1  & 4.9     & 33      & 557   & 85722            \\ \hline
2  & 5.1     & 35      & 614   & 96180            \\ \hline
3  & 5.3     & 35      & 627   & 98260            \\ \hline
4  & 6.0     & 40      & 739   & 124284           \\ \hline
5  & 6.2     & 41      & 802   & 136518           \\ \hline
\end{tabular}
\label{tab:proguard}
\end{table}

\subsection{Metrics Analysis}
After Collecting the projects, the source code is parsed using JavaParser \cite{hosseini2013javaparser}. It provides the Abstract Syntax Tree (AST) of the code of the input system. After that, the metrics values are calculated from the parsed AST. From the metrics values, the metrics information is collected. After that, the provided metrics as mentioned in Section \ref{sec:introduction}, are stored and the necessary metrics are mapped with the illustration metaphors as described in Section \ref{sec:tool}.

\subsection{Software Visualization}
JavaFX \cite{ebbers2014mastering} has been used for handling the GUI of the system. It is an open-source platform built-in java language. Based on the metrics values, all the graphic elements are created using the library JavaFX. For making the map interactive, there are keyboard and mouse listeners to explore the UI.

\begin{figure}[ht]
\centering
    \begin{subfigure}[b]{0.33\textwidth}
    \includegraphics[width=\textwidth]{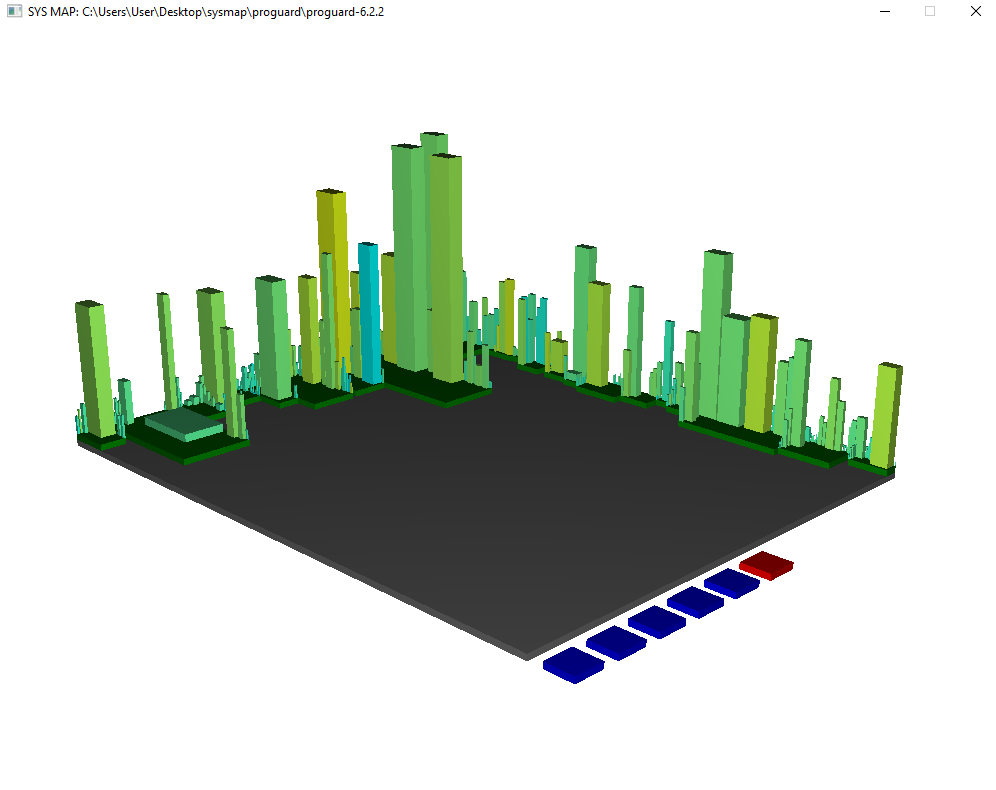}
    \caption{Normal View}
    \label{fig:o1}
    \end{subfigure}
    ~
    \begin{subfigure}[b]{0.33\textwidth}
    \includegraphics[width=\textwidth]{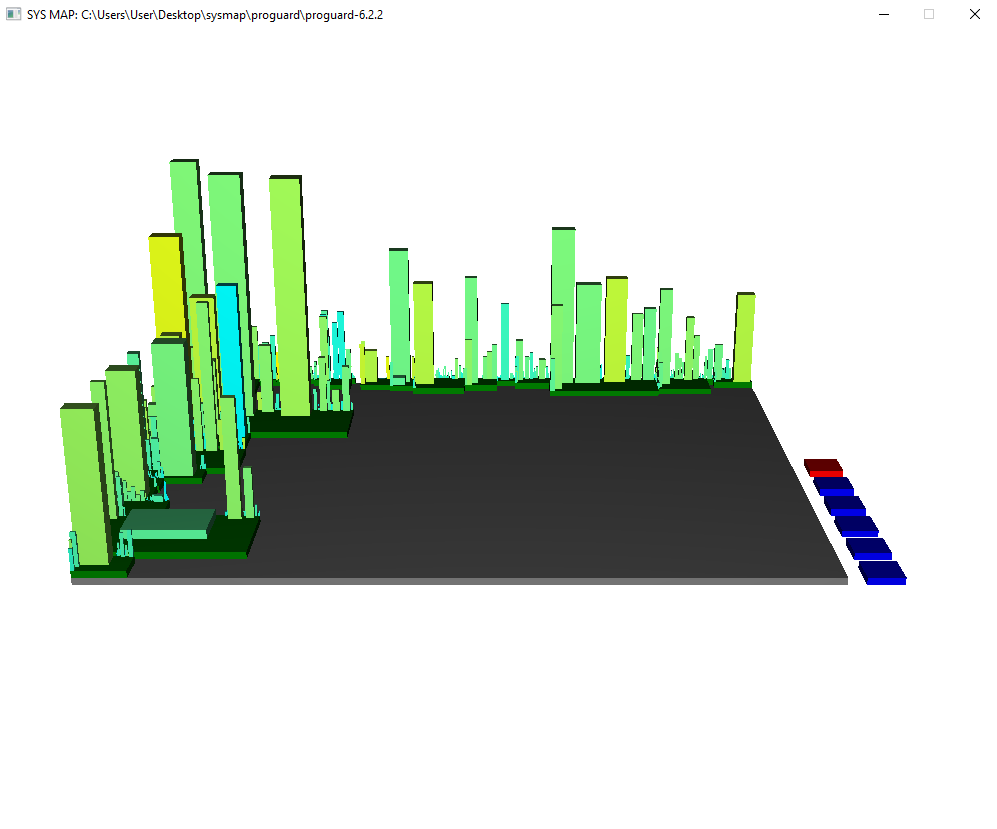}
    \caption{Side View}
    \label{fig:o2}
    \end{subfigure}
    ~
    \begin{subfigure}[b]{0.33\textwidth}
    \includegraphics[width=\textwidth]{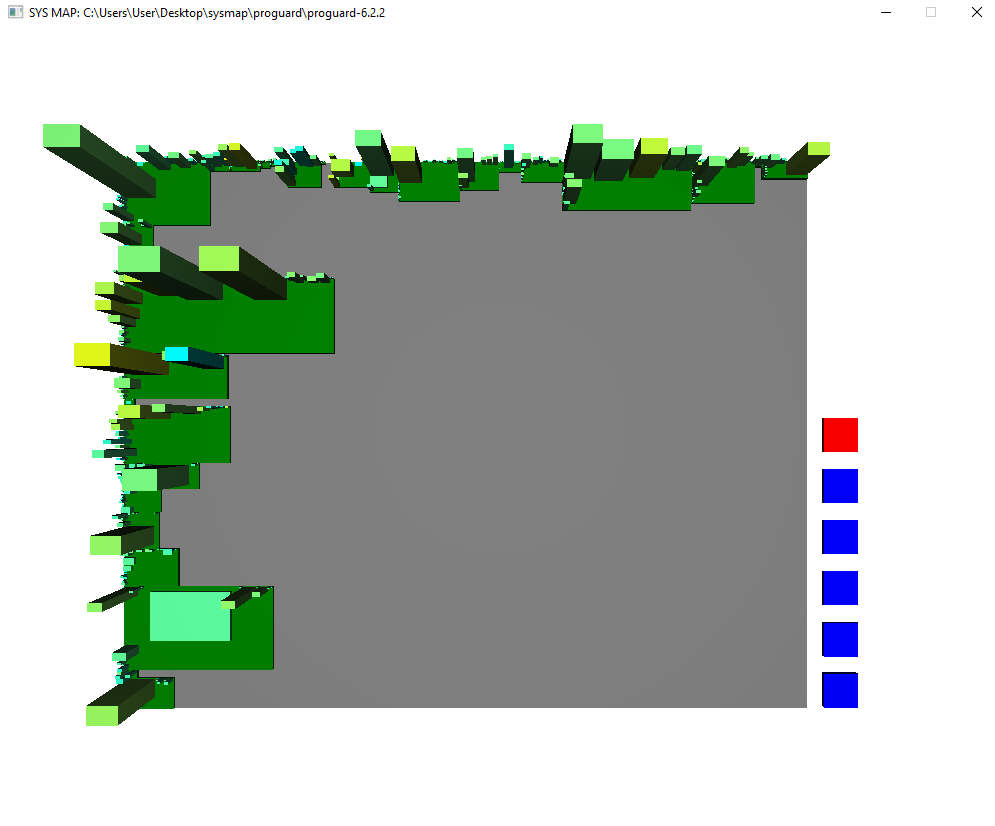}
    \caption{Top View}
    \label{fig:03}
    \end{subfigure}
    ~
    \begin{subfigure}[b]{0.33\textwidth}
    \includegraphics[width=\textwidth]{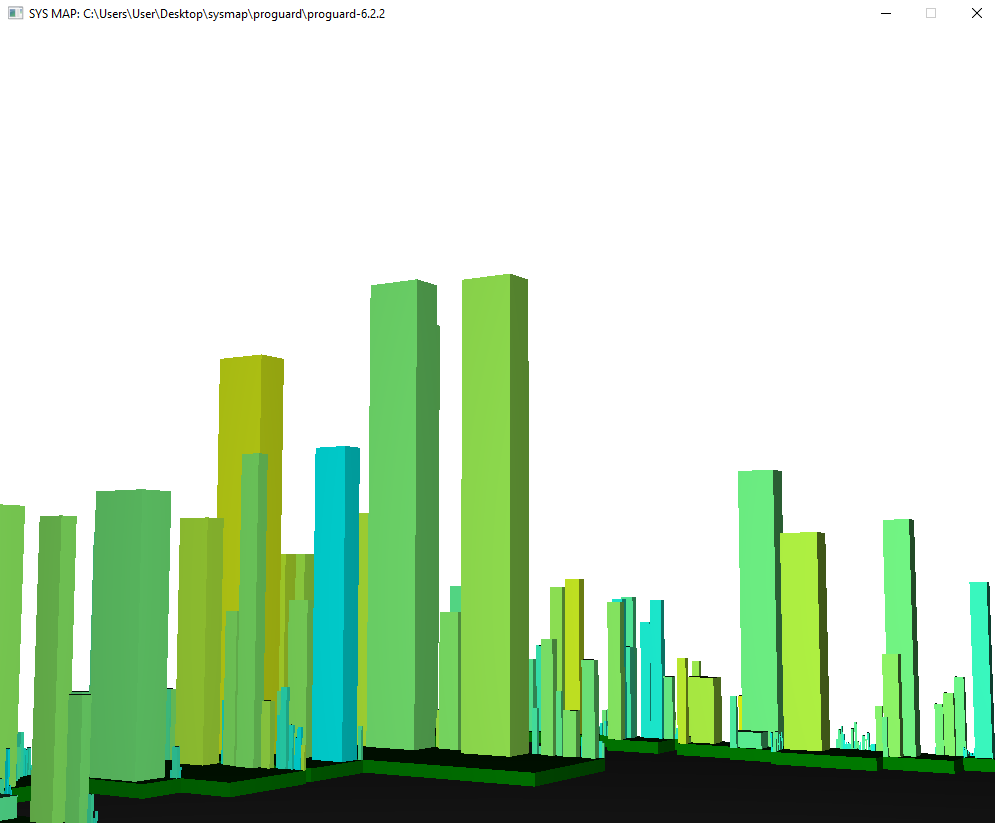}
    \caption{Close View}
    \label{fig:04}
    \end{subfigure}
\caption{Different View of the Map of a System}
\label{fig:overview}
\end{figure}


\subsection{Evolution Analysis}
From the calculated metrics values of different versions of a system, the data is collected for analyzing evolution. For the time being, the metrics are considered for evolution calculation are - no of packages, no of classes, LOC and WMC. These values are considered because this tool tries to simplify the overall illustration and comprehend the evolution in minimal consideration.

For denoting the complex class, the term skyscrapers  has been used. To identify the skyscrapers  limit, a simple Equation \ref{eq:ssl} is used. All the buildings that crosses the limit are identified as skyscrapers , i.e. complex classes.
    \begin{equation}
    \label{eq:ssl}
        skyscrapers  height limit =  2 \times \frac{1}{n} \sum_{i = 1}^n (WMC)
    \end{equation}
    
As for the god class, the term heavy building has been used. To identify the limit for heavy building, another Equation \ref{eq:hcl} is used. All the buildings whose base crosses the limit are identified as heavy buildings, i.e. god classes.
    \begin{equation}
    \label{eq:hcl}
        Heavy building base limit =  2 \times \frac{1}{n} \sum_{i = 1}^n (LOC)
    \end{equation}
    
In the equations, 

    \textit{n} = Total number of classes
    
    \textit{WMC} = Weighted method per class
    
    \textit{LOC} = Line of Code
    
The metrics are plotted in a bar chat to illustrate the version to version change. JFreeChart \cite{gilbert2002jfreechart} is used to draw the charts. 

\section{Tool Description}
\label{sec:tool}
There are several tools that help in 3D visualization of a full software system. However, in the Section \ref{sec:rel}, the shortcomings for previous solutions are pointed out. One of the main intentions of this approach is to make a lightweight tool where users can easily interact to extract the visual information regarding software evolution.


\subsection{SysMap Metaphor}
The elements of the full illustration indicate different properties of the software system. As it is shown in \ref{fig:overview}, several views type such as-normal view, side view and top view are introduced for better visualization.

\textbf{Area Boundary:} The area of the output illustration represents the vastness of the system. The color of the land area is grey.

\textbf{Plots:} The green-colored plots represent the packages of the software. These plots may hold several buildings in the area. The size of the plot depends on the size of the buildings it holds.

\textbf{Buildings:} The buildings are the representation of the classes. The base of a building is the line of code of that class. As for the height, the weighted method of a class is taken into account. Line of Code and weighted method of a class are chosen for the illustration because these two metrics can show how much a class is big and complex respectively. The color of the buildings varies because those represent the coupling of that class with others. The more the color is yellowish, the more the value of the coupling is and the color will be cyan otherwise. The interfaces are represented just like the classes. To avoid heavy graphics, only these metrics are used for minimal illustration.


\begin{figure}
    \centering
        \begin{subfigure}[b]{0.32\textwidth}
        \centering
        \includegraphics[width=\textwidth]{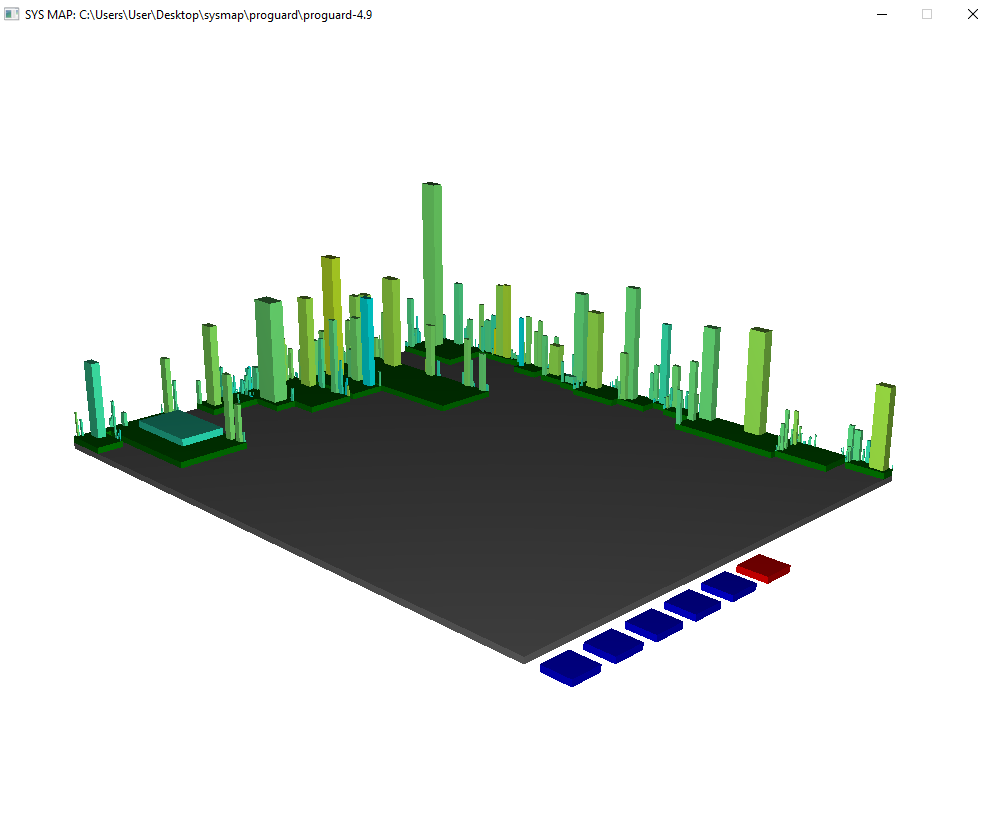}
        \caption{Proguard v-4.9}
        \label{fig:p4.9}
    \end{subfigure}
    \hfill
    \begin{subfigure}[b]{0.32\textwidth}
        \centering
        \includegraphics[width=\textwidth]{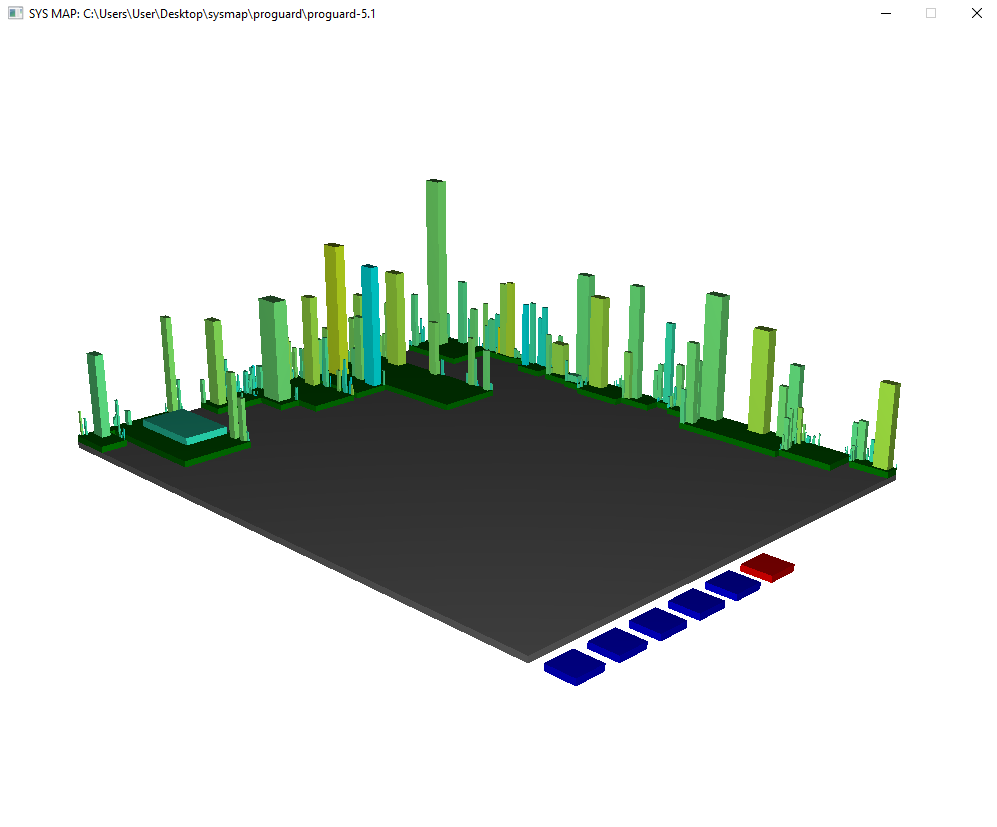}
        \caption{Proguard v-5.1}
        \label{fig:p5.1}
    \end{subfigure}
    \hfill
    \begin{subfigure}[b]{0.32\textwidth}
        \centering
        \includegraphics[width=\textwidth]{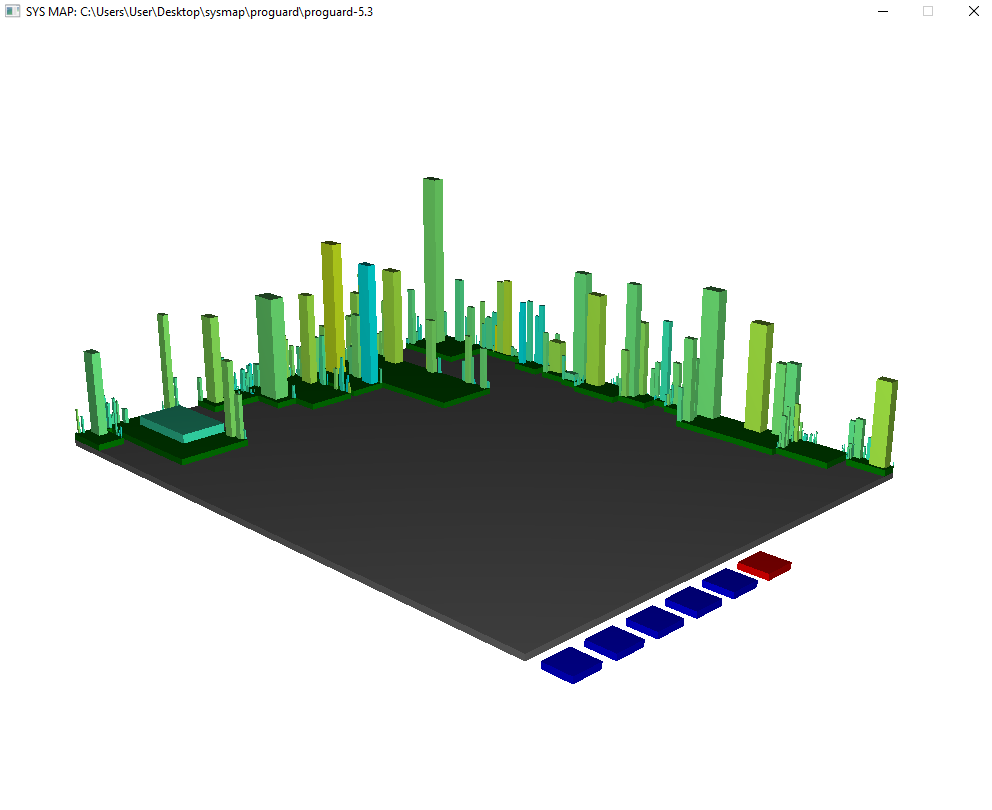}
        \caption{Proguard v-5.3}
        \label{fig:p5.3}
    \end{subfigure}
    \begin{subfigure}[b]{0.45\textwidth}
        \centering
        \includegraphics[width=\textwidth]{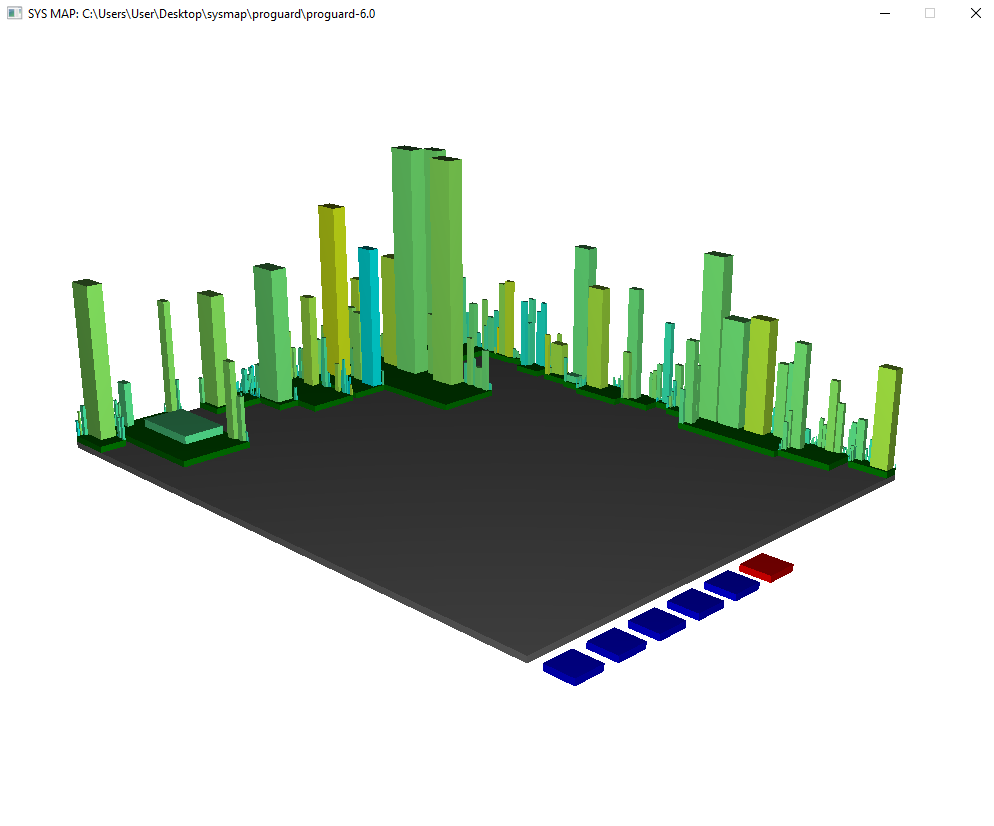}
        \caption{Proguard v-6.0}
        \label{fig:p6.0}
    \end{subfigure}
    \hfill
    \begin{subfigure}[b]{0.45\textwidth}
        \centering
        \includegraphics[width=\textwidth]{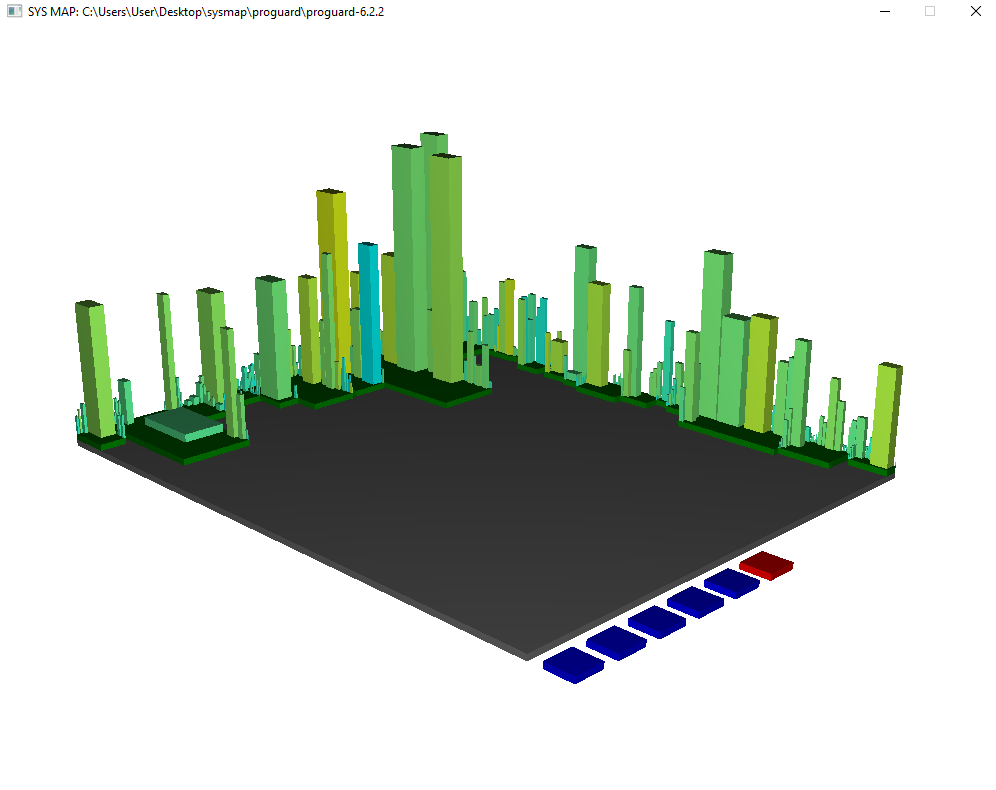}
        \caption{Proguard v-6.2.2}
        \label{fig:p6.2.2}
    \end{subfigure}
    \caption{Maps of Proguard through Different Versions}
    \label{fig:evol}
\end{figure}

\begin{table}[ht]
\centering
\caption{Keyboard Key and Actions}
\begin{tabular}{cc|cc}
\hline
\textbf{Key} & \textbf{Action} & \textbf{Key} & \textbf{Action}  \\ \hline\hline
Up  & Look Above      & A            & Move Left        \\
Down         & Look Down       & D            & Move Right       \\
Left         & Look Left       & Q            & Move Above       \\
Right        & Look Right      & E            & Move Below       \\
W            & Move Forward    & S            & Move Backward   \\    \hline
\end{tabular}
\label{tab:keyAction}
\end{table}

\subsection{Interaction with UI}
To make the illustration more comprehensible several interaction options have included in SysMap. The user can easily explore the map through these options. The interaction options are:

\textbf{Navigation:} The user can have different views of the map, such as - front view, side view, top view, back view. These views can be obtained by changing the camera angle of this 3D visualization. Also, to have a close look, zoom-in option is available and for the overview of the system one can zoom out the map. The keyboard listeners are used to handle these actions. In Table \ref{tab:keyAction}, all the actions corresponding to the keys are mentioned

\begin{figure}[ht]
  \centering
  \includegraphics[width=.7\textwidth]{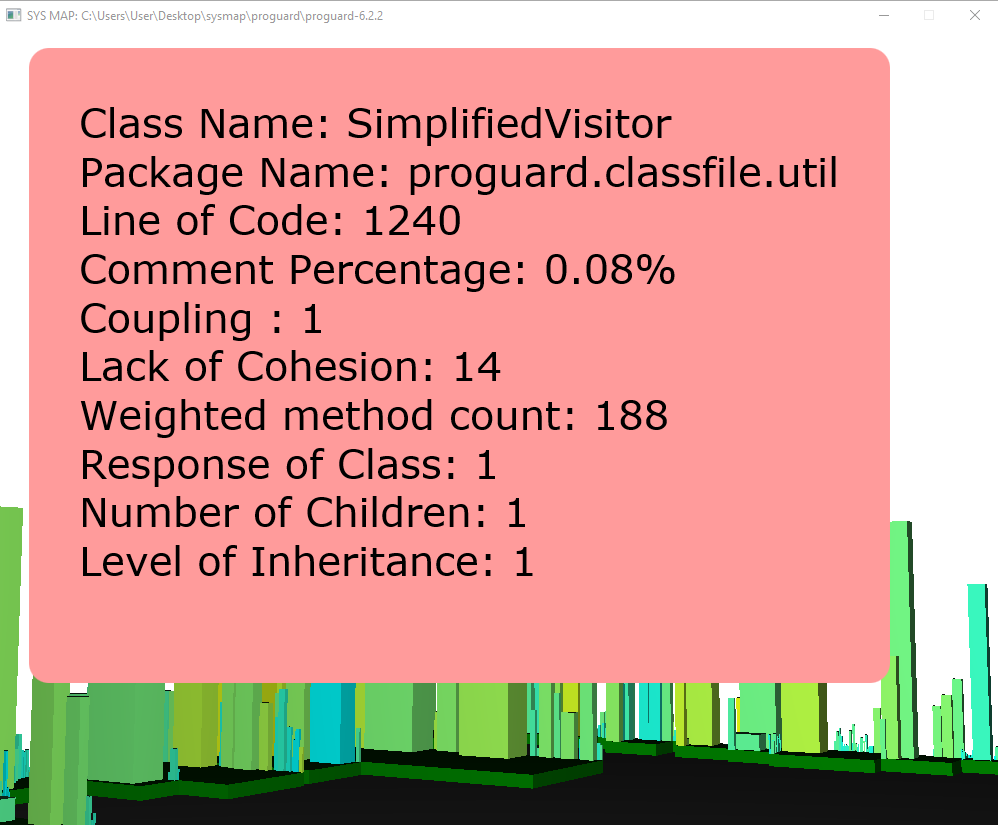}  
  \caption{Metrics Information Shown after Clicking a Building}
  \label{fig:popUp}
\end{figure}

\textbf{Element Selection:} The users can have a quick glance at some of the metrics of the classes and packages by clicking the plots and the buildings. For clicking a building, a message box pops up to provide related information represented by the element as shown in Figure\ref{fig:popUp}. 

\textbf{Version Traversing:} Beside the map of the city, there will be some blue colored buttons according to the number of the versions. By selecting the buttons, the user can view the map of that respective version. So, this option helps a user to have a look at several versions at a time.

\textbf{Evolution Checker:} There is a red button along with the version changer buttons that helps to check the evolution of the system. After clicking this button, a bar chart is popped up. The chart is prepared on the basis of four previously mentioned metrics of different versions. An video demonstration of SysMap is available in the Google drive \cite{drive}.


\section{Result Analysis}
\label{sec:res}
Different 3D visualization tools use different strategies to provide more usability. To analyze the differences the output illustration of SysMap and CodeCity is compared. Also, the evolution of a software system can be clearly described using this tool. Even for a significant change, the developers can easily track down the classes right from the illustration.

\subsection{System Visualization}

Since different metrics are used as metaphors to build the city of CodeCity than SysMap, the visual representation of the same system is very different. In CodeCity, the base of a building is number of attributes and the height is number of methods where in SysMap, these are LOC and WMC respectively. So, the heavy buildings and the skyscrapers  are also incomparable in such scenario.

\subsection{System Evolution}
Based on the information showed in Table \ref{tab:proguard}, it is clearly understandable that how a project is evolved over the year. However, this requires intensive analysis to get the changes. This total work process is simplified by the proposed tool as one can easily visualize the high level modification from the provided illustration. in Figure \ref{fig:evol}, five versions of the Proguard system is shown. With the change of version, many buildings got taller or wider. In the right side of the fourth map some buildings are missing which were seen in the third map. Some buildings are more yellowish or bluer in the latest versions. Therefore, all these indicates how the values of these metrics changed over the time.

\begin{figure}[ht]
  \centering
  \includegraphics[width=.8\textwidth]{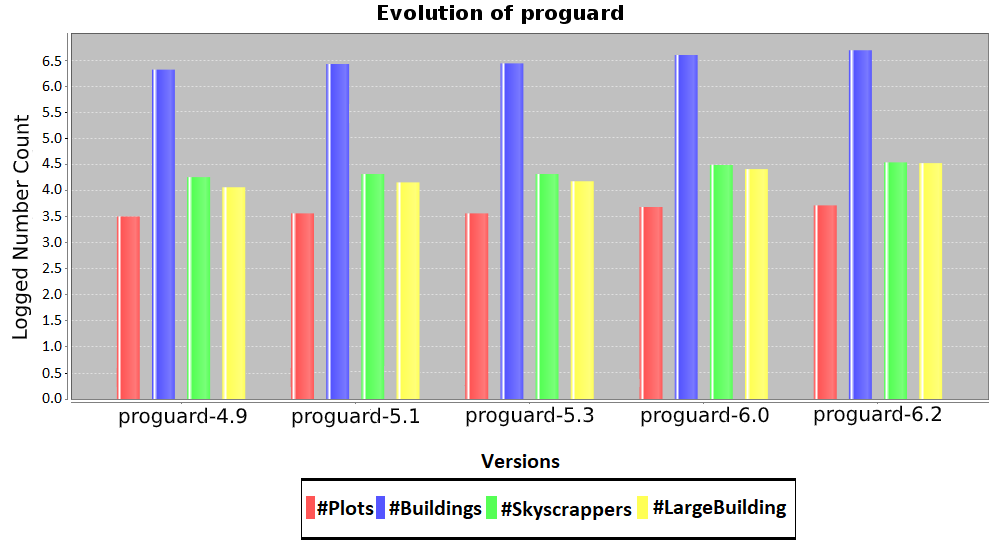}  
  \caption{Evolution of Proguard through different versions}
  \label{fig:chart}
\end{figure}

After processing the source code some values of the metrics are found to build the chart for analyzing system evolution. In Figure \ref{fig:chart}, it is seen that, in the x-axis, different versions along with their four metrics are denoted, where in y-axis, the \textit{e} based log values are marked.It is clearly visible that, with the time all the values of the metrics increased.


\section{Threats to Validity}
\label{sec:threat}
The outcome of this study may have some internal and external threats to validity.The procedure of computing the metrics values and setting several threshold values may cause the internal validity issues.The external validity threats can be explained by the study subjects and the way of evaluating the output results.

\textbf{Metrics Calculation:} The metrics calculation rule or formula may be a reason to have a different view of a system. Some of these metrics' calculations may vary occasionally. So, this issue is considered to be a threat.

\textbf{Threshold Values:} This approach needed some threshold value during handling the GUI material. Such as - a minimum threshold value is needed to be set to LOC far a class to be a building material. This helps to focus on bigger and complex classes. Such a decision may cause some variation to the maps of same systems.

\textbf{Validation Metrics:} The selected metrics and the method of trimming their values are selected for the convenience of result analysis. Different measurement may cause different outcome.

\textbf{Manual Inspection:} The validity of the metrics values and output illustrations are checked manually. Two different developers cross verified the elements with the source code. Such manual inspection also considered as a threat to the validity of this system.

\textbf{Study Subjects:} The systems which are examined to evaluate the proposed approach are collected from the open-source repositories. These systems may have syntax error parsing issues, bugs and missing information.


\section{Related Works}
\label{sec:rel}
\noindent A lot of works on software visualization has been done in the early decade of this century. But at the present time, 3D visualization is more popular and comprehensible than 2D visualization. Also, while some visualization methods are providing for the illustration of the full system using some metaphor, there are some approaches which rather show the illustration of the code flow.

To understand the structural evolution of a system a technique was proposed \cite{telea2008code} that captures the graphical representation of the code flows. Using this, one can detect the code drifts, splits, merges, insertions, deletions and so on. The major problem of such a technique is, for a large and complex class, this illustration gets very complex which is not comprehensible for humans. Besides, this approach is somewhat code specific rather than project specific.

A solar system metaphor was proposed \cite{graham2004solar} where the planets represent java classes, orbits are for inheritance level and planet sizes depend on the Line of codes (LOC) of the classes. This method tried to capture the inheritance, coupling, and cohesion among the classes. However, this method does not have much relevance and cannot help in illustrations for the complex systems properly \cite{wettel2007visualizing}.

There are several approaches that have used the city metaphor to present the graphical view. In paper \cite{wettel2007visualizing}, a tool built in Smalltalk language named CodeCity  was proposed. There the buildings represent the class files in a system, where the base of the buildings are the number of attributes and the height of the buildings are the total number of the methods. They extended their work by considering the granular level in \cite{wettel2008visual}. Evospaces are another tool proposed in \cite{alam2007evospaces}. They put some fixed types of buildings based on the LOC range of the class files and showed people inside the buildings which represent the methods. Another tool named CodeMetropolis \cite{balogh2013codemetropolis} was proposed, which is a kind of similar approach. There the formation of the buildings is based on the code and comments of the codes. They proposed another version of CodeMetropolis \cite{balogh2015codemetropolis} where the illustration is interactive like a video game. This tool can capture every detail of the systems since they have used the game engine. However, these features make the tool very heavy for the machine. Another tool named Code Park is proposed \cite{khaloo2017code}, where the classes are represented in smaller rooms. The room contains the codes of the classes in its wall so that the user can check those out. But the graphical representation is quite like the IDEs and not very helpful to understand the overview.

The previously mentioned tools have not illustrated the software evolution. Though some of them \cite{wettel2007visualizing}, \cite{wettel2008visual}  theoretically explained the changes of different releases, but no automated solution was provided. The relationship between software metrics and evolution is analyzed in \cite{lehman1997metrics}, \cite{drouin2013analyzing}. In \cite{mens2001evolution}, the software metrics are analyzed to find the contributing metrics for software evolution. The size metrics i.e. SLOC and number of files are analyzed to find their role in evolution in \cite{herraiz2006comparison}. But these papers did not provide any graphical solution. In paper \cite{lanza2002understanding}, they tried to show the software evolution by using simple graphical view and metrics. But those illustrations are not much comprehensible and hold less information.


\section{Conclusion}
\label{sec:con}
In this paper, a tool named SysMap was proposed that provided a 3D visualization to illustrate a software system. The total area represents the full system, the plots are its packages and the buildings are the classes. Eight code metrics were calculated and depending on LOC, WMC and coupling between classes each of the elements was made. This system only focused on the evolution of java projects which maintain the object-oriented concepts. Several large open-source projects and small student projects were analyzed to understand the efficiency of the tool.

The solution of SysMap provided a very minimal graphics to make it a lightweight tool. The user can easily interact with the tool through any regular machine because it would not require any high computation. The elements of the map provided metrics related information. This paper provided the evolution information of five versions of an open-source project named Proguard. Using a bar chart, the evolution was shown in the UI of the tool.

The future work of this tool will aim to incorporated other metrics to illustrate the elements differently. Also, for analyzing the software evolution, more metrics and a flexible threshold will be considered in SysMap.


\bibliographystyle{unsrt}  
\bibliography{sysmap}  

\end{document}